\documentclass{webofc}
\usepackage[varg]{txfonts}   
\usepackage{wrapfig}
\newcommand \BQ {\chi_{11}^{BQ}}
\newcommand \BS {\chi_{11}^{BS}}
\begin{document}
\title{Conserved charge fluctuations at vanishing net-baryon density from Lattice QCD}
%
%

\author{\firstname{Jishnu} \lastname{Goswami}\inst{1}\fnsep\thanks{\email{jishnu@physik.uni-bielefeld.de}} \and
        \firstname{Frithjof} \lastname{Karsch}\inst{1}\and
        \firstname{Swagato} \lastname{Mukherjee}\inst{2}\and
          \firstname{ Peter} \lastname{Petreczky}\inst{2}\and
         \firstname{Christian} \lastname{Schmidt}\inst{1}
}

\institute{Fakult\"at f\"ur Physik, Universit\"at Bielefeld, D-33615 Bielefeld,
		Germany
\and
           Physics Department, Brookhaven National Laboratory, Upton, NY 11973, USA
          }

\abstract{%
 We present here continuum extrapolated results for all $2^{nd}$ order
 cumulants using the most
 resent results obtained by the HotQCD collaboration in (2+1)-flavor QCD.
  We constrain the applicability  of various HRG models by presenting a detailed comparison of our results based on different sets of hadron spectra as well as with virial expansion based model calculations. A
 comparison with our lattice QCD results for $2^{nd}$ order cumulants with
 models that parametrize repulsive interactions among baryons and anti-baryons
 in a hadron resonance gas through a single excluded volume parameter (EVHRG) is also shown.
 }
\maketitle
\section{Introduction}
\label{intro}
The partition function of quantum chromodynamics (QCD) depends on the temperature (T) and the chemical potentials, $\mu_B$, $\mu_Q$ and $\mu_S$ corresponding
to three conserved charges namely baryon number (B), electric charge (Q) and strangeness (S). Fluctuations of these conserved charges can be calculated from first principles Lattice QCD formalism. On the other hand two large experimental programs at RHIC in USA and LHC in Switzerland measure the fluctuations of conserved charges either from event by event collisions or from particle yields. 
In those experimental programs the temperature and chemical potential at the time of freeze out is determined by using hadron resonance gas (HRG) models. Therefore it is essential to constrain the applicability of HRG models by a detailed comparison with lattice QCD, which is the fundamental theory of strong interactions~\cite{review}.

In these proceedings we present all the $2^{nd}$ order cumulants calculated by the HotQCD collaboration at vanishing chemical potential in (2+1) flavor QCD and show a detailed comparison with various HRG models. We show that when comparing HRG models with QCD the details of the spectrum play an important role for observables dominated by strange particles. For instance, a non interacting point like HRG model with established resonances from tables provided by the particle data group (PDG) (3 and 4 star states)~\cite{Zyla:2020zbs} works poorly in the strangeness sector. Thus to improve the baryon-strangeness correlations  additional resonances from PDG tables (1 and 2 star) and quark models\cite{Isgur,Ebert} are required. Baryon-electric charge correlations show some striking robust feature close to the pseudo-critical transition temperature of QCD which we will discuss in detail in the following section.

We also show results from model calculations that include additional interactions through excluded volume terms (EVHRG) and a virial expansion based formalism, respectively.  

\begin{figure}[t]
	\centering
	\includegraphics[width=5.1cm,clip]{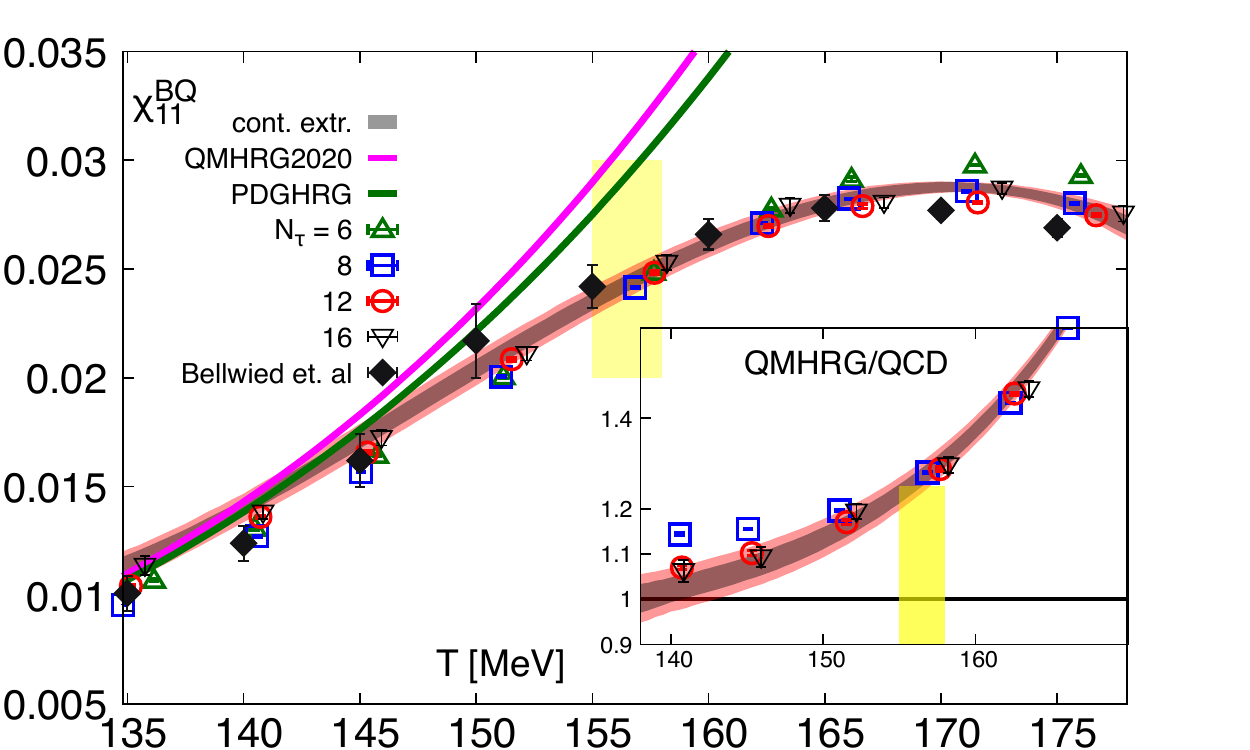}
	\includegraphics[width=5.1cm,clip]{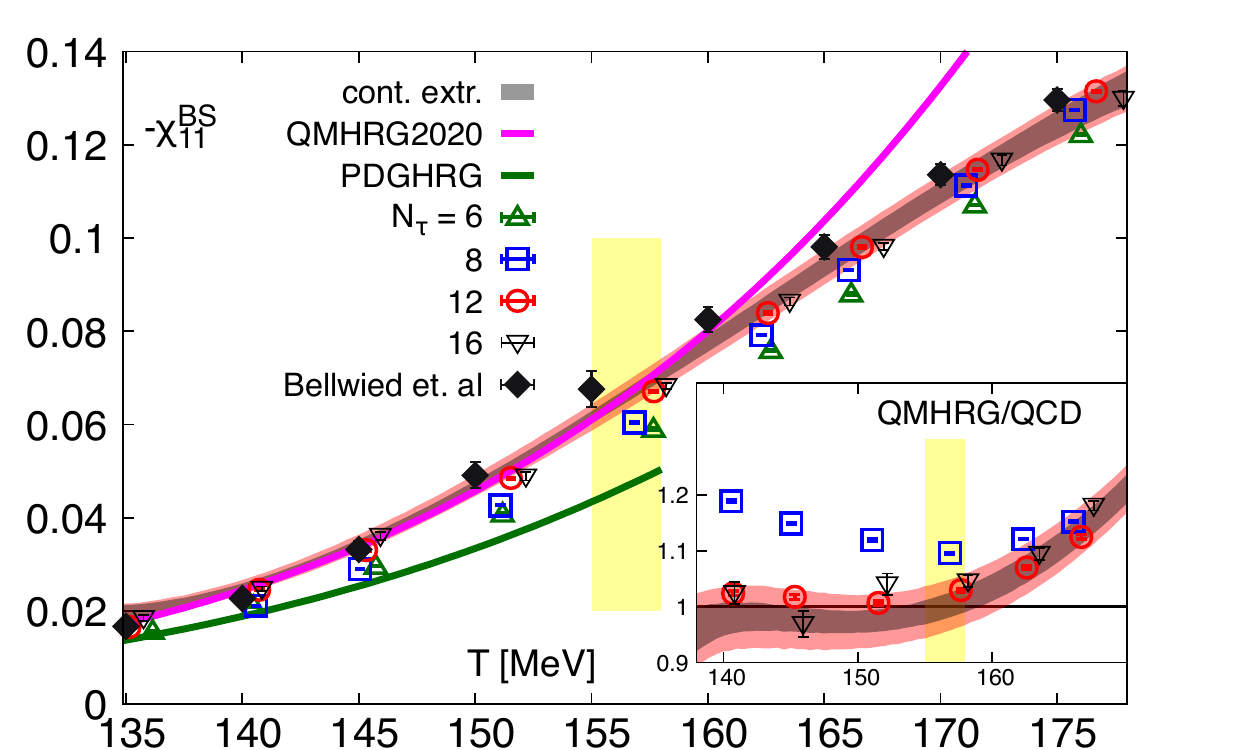}
	\includegraphics[width=5.1cm,clip]{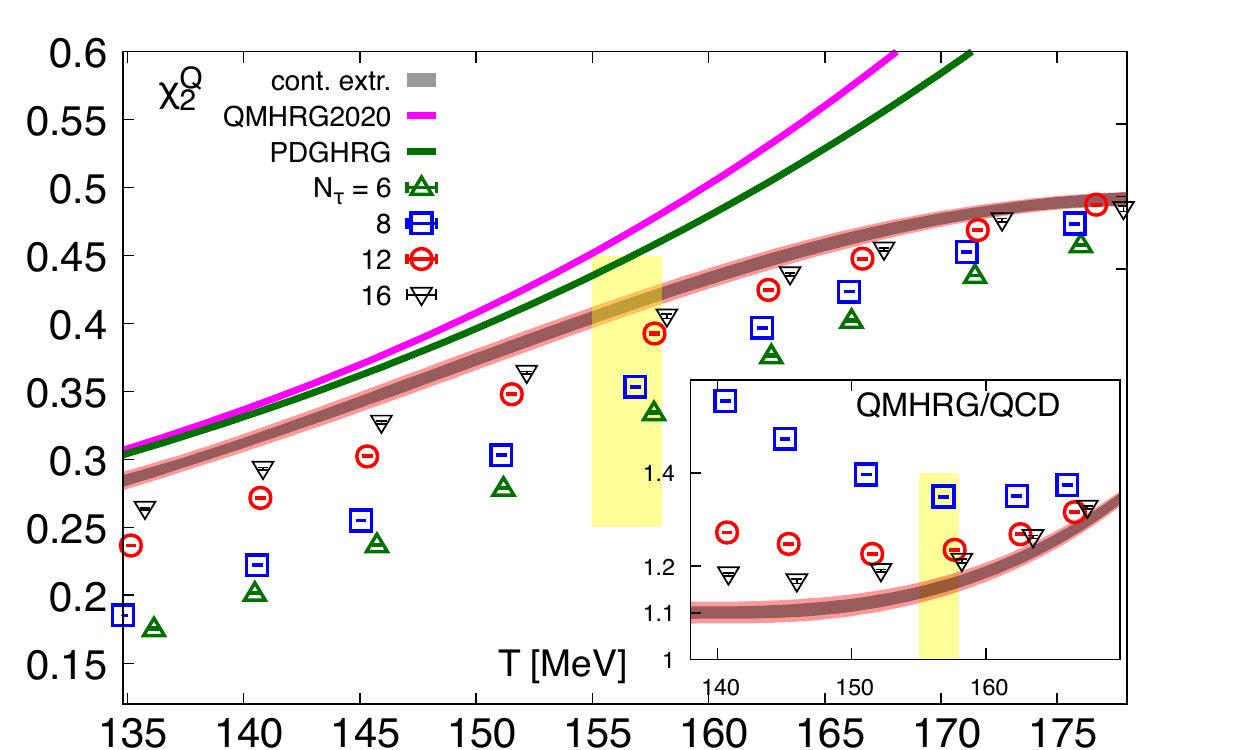}
	\includegraphics[width=5.1cm,clip]{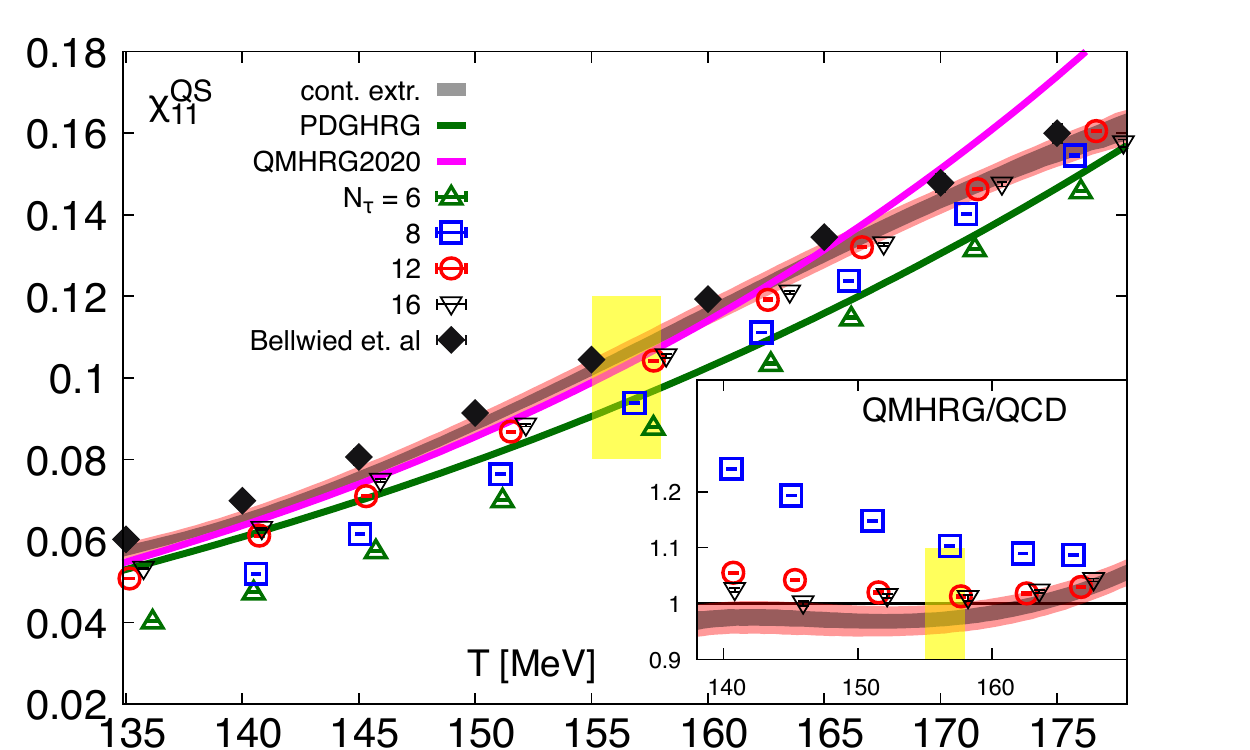}
	\caption{Four independent $2^{nd}$ order cumulants versus temperature calculated
		on lattices with different temporal extent $N_\tau$. Lines show results from
		HRG model calculations using point-like, non-interacting resonances 
		and the hadron spectrum list QMHRG2020
		\cite{Bollweg:2021vqf}. 
		Also shown are results from Bellwied et al. \cite{Bellwied:2019pxh}.
		The insets show the ratio of continuum extrapolated lattice QCD results
		and HRG model calculations based on QMHRG2020.
		The yellow band corresponds to the crossover temperature, $T_{pc}$.}
	\label{fig-1}       
\end{figure}

\section{\boldmath$2^{nd}$ order cumulants and QMHRG2020}
\label{sec-1}
In our calculations we have used highly improved staggered quarks (HISQ) with two light (u,d) and one strange (s) flavors. The pressure can be written as
\begin{eqnarray}
	P/T^4 = \frac{1}{VT^3} \ln Z(T,V,\vec{\mu}),
\end{eqnarray}
where $\vec{\mu} = (\mu_u,\mu_d,\mu_s)$ and $Z$ is the QCD partition function,
\begin{eqnarray}
	Z=\int\mathcal{D}U\;\text{det}[M(m_u,\mu_u)]^{1/4}\text{det}[M(m_d,\mu_d)]^{1/4}\;  \text{det}[M(m_s,\mu_s)]^{1/4}\;e^{-S_G(U)}\
\end{eqnarray}
The fluctuations and correlations at vanishing chemical potentials can be obtained as
\begin{eqnarray}
\chi_{lmn}^{BQS} = \frac{\partial P^{l+m+n}/T^4}{\partial (\mu_B/T)^l \partial (\mu_Q/T)^m \partial (\mu_S/T)^n}\bigg|_{\vec{\mu}=0} ~,
\end{eqnarray}
where 
derivatives with respect to conserved charge chemical potentials, $(\mu_B,\mu_Q,\mu_S)$, have been expressed
in terms of the flavor chemical potentials $\vec{\mu}$~\cite{HotQCD:2012fhj}. 
Since in (2+1)-lattice QCD calculations one uses degenerate light quark masses, $m_u=m_d$,
the diagonal susceptibilities, $\chi_2^B$ and
$\chi_2^S$ can be obtained using the relations
$	\chi_2^S = 2 \chi_{11}^{QS} - \chi_{11}^{BS}$
and	$\chi_2^B = 2 \chi_{11}^{BQ} - \chi_{11}^{BS}$\; .
 
 In Fig.(\ref{fig-1}) we present continuum extrapolations for four independent $2^{nd}$ order cumulants, and a comparison with other lattice calculations is also shown~\cite{Bellwied:2019pxh}. In HRG model calculations with point like non-interacting resonance the non-strange baryons, strange baryons, pions and kaons provide the major contributions to $\chi_{11}^{BQ}$ ,  $\chi_{11}^{BS}$ , $\chi_{2}^{Q}$ and $\chi_{11}^{QS}$ respectively. We compare these observables with HRG model calculations based on only established resonances (PDGHRG) and it is clear that for all the $2^{nd}$ order observables PDGHRG works poorly at $T>140~\rm{MeV}$. We define QMHRG2020 by including less established resonances from PDG tables 
 (1-star and 2-star)  and additional resonances from Quark Model (QM) calculations. In \cite{Bollweg:2021vqf}  QMHRG2020 is discussed in great detail. QMHRG2020 describes the  lattice data of $\chi_{11}^{BS}$  and $\chi_{11}^{QS}$ quantitatively better than 1\% in the temperature range $T < 156~\rm{MeV}$ indicating the abundance of strange particles in nature. However, at the same time the additional non-strange baryons included in QMHRG2020 have little effect on $\BQ$ correlations. Although, QMHRG2020 provides better than 5\% level agreement in the temperature range $T \leq 150~\rm{MeV}$, it still exhibits an almost 20\% deviation close to the pseudo-critical temperature, $T_{pc}$. Since $T_{pc}$ \cite{Bazavov:2018mes} is close to the freeze-out temperature~\cite{Stachel:2013zma} measured by the ALICE collaboration, quantifying the agreement/deviation between HRG model calculations and QCD close to $T_{pc}$ is the most relevant for understanding the thermal properties of the dense, strongly interacting medium created in heavy ion collision experiments. 
 
 Finally we note that electric charge fluctuations, $\chi_{2}^Q$, deviate about 10\% close to $T_{pc}$
 from HRG model calculations (see \cite{Bollweg:2021vqf} for further discussion).
 Also the temperature derivatives of all cumulants 
 calculated in HRG models are quite different from QCD at $T>156~\rm{MeV}$ indicating that a hadronic description of the thermal medium does not seem
 to be appropriate for temperatures beyond $T_{pc}$.


\begin{figure}
	\centering
	\includegraphics[width=5.6cm,clip]{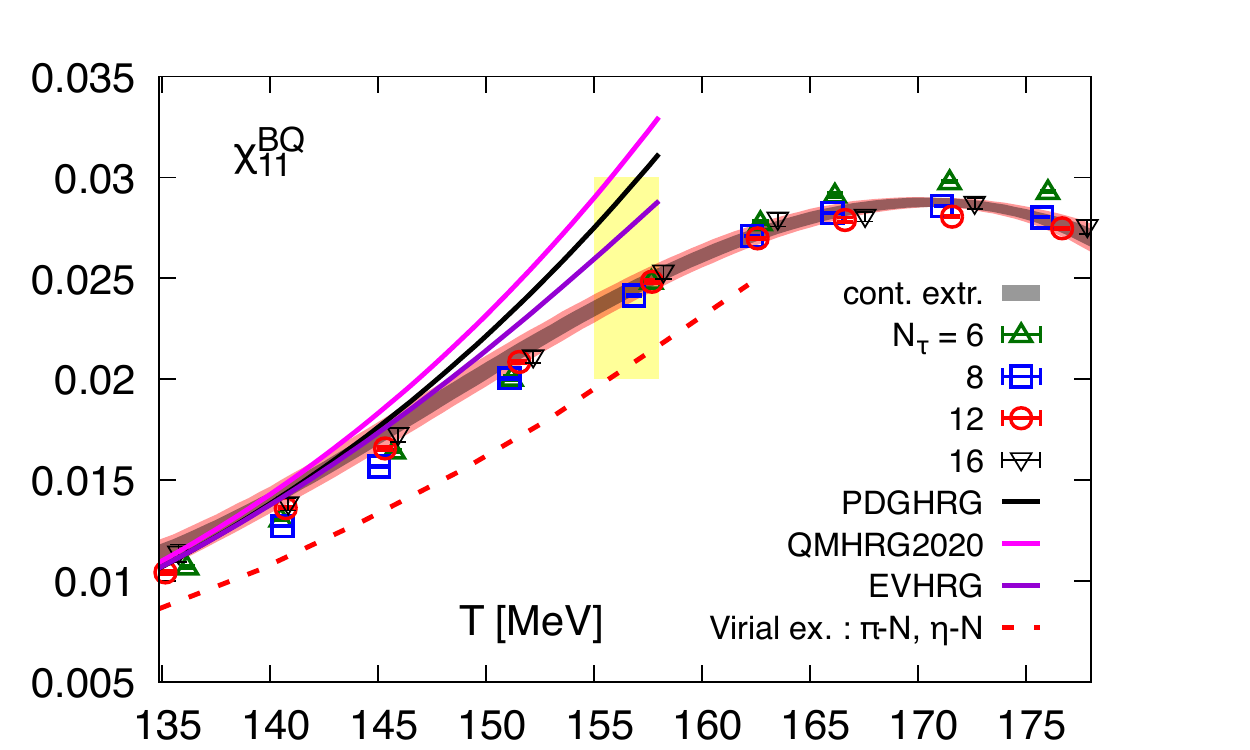}
	\includegraphics[width=5.6cm,clip]{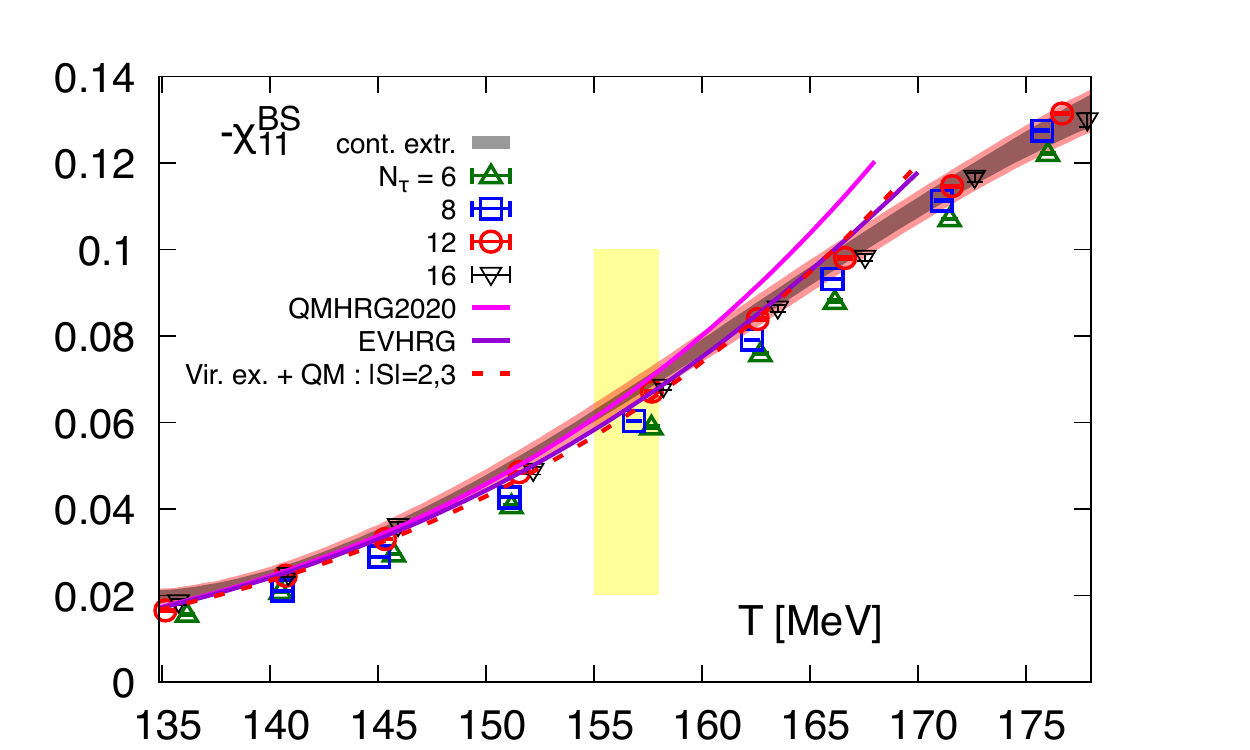}
	\caption{Continuum extrapolated results for 
		$\chi_{11}^{BS}$ (left) and $\chi_{11}^{BQ}$ (right).
		Shown is a comparison with HRG model calculations
		based on QMHRG2020 as 
		discussed in the text. Also shown are results obtained
		with excluded volume HRG models, using an excluded volume parameter $b=1$~fm$^3$, and $2^{nd}$ order virial expansions
		\cite{Lo:2017lym,Fernandez-Ramirez:2018vzu},
		respectively.}
	\label{fig-2}       
\end{figure}
\section{Net baryon-electric charge and baryon-strangeness fluctuations at \boldmath$T_{pc}$}
Preliminary results for  $\chi_{11}^{BQ}$  and  $\chi_{11}^{BS}$ close to the $T_{pc}$ have been presented by us already in ~\cite{Goswami:2020yez}. In Fig.(\ref{fig-2}) (left), excluded volume HRG (EVHRG) calculations considering a hard sphere radius of $b=1~\rm{fm}^3$~\cite{Vovchenko:2016rkn,Karthein:2021cmb} are shown. It is evident that one needs $b> 1~\rm{fm}^3$  for a proper description of $\BQ$. This, however is in conflict with the nice agreement 
obtained for $\BS$. Thus one needs at least two distinct $b$ values for the non-strange baryons and strange baryons for a better agreement with lattice data. Furthermore, we also show the $\BQ$ correlations calculated from $2^{nd}$ virial coefficient \cite{Lo:2017lym}, which falls below the Lattice data. For a better quantitative comparison, one needs information from higher order virial coefficients which is not known but has been modeled using LQCD calculations~\cite{Andronic:2018qqt}. It is not shown here. The virial expansion based calculations of $\BS$ works better but currently cannot treat multi-strange baryons which in Fig.(\ref{fig-2}) have been added as a non-interacting QMHRG contributions \cite{Fernandez-Ramirez:2018vzu}. Since, $\chi_2^Q$ and  $\chi_{11}^{QS}$ are dominated by pions and kaons respectively, EVHRG doesn't play any significant role here. 

\section{Conclusions}
The qualitative features of $2^{nd}$ order cumulants of conserved charge fluctuations and correlations, calculated in lattice QCD, are reasonably well described by point like non-interacting HRG models up to the pseudo-critical temperature for the QCD transition. We also present calculation based on $2^{nd}$ order virial expansions. We specify that for a better quantitative agreement of $\BQ$ with Lattice QCD results higher order expansions are needed.
We furthermore compared our lattice QCD results for $2^{nd}$ order cumulants with
model calculations that parametrize repulsive interactions among baryons and anti-baryons
in a hadron resonance gas through a single excluded volume parameter (EVHRG).
We point out that such an approach is not sufficient to describe all $2^{nd}$
order cumulants simultaneously. At least independent excluded volume parameter
for strange and non-strange baryons would be needed already for the description of $2^{nd}$ order cumulants. Furthermore, all the HRG-based calculations seem to fail to describe lattice results for $T >T_{pc}$ already at the level of $2^{nd}$ order cumulants.

\section*{Acknowledgements}
This work was supported by the Deutsche Forschungsgemeinschaft (DFG) - project number 315477589 - TRR 211, the European Union H2020-MSCA-ITN-2018-813942 (EuroPLEx), and the U.S. Department of Energy, Office of Science, Office of Nuclear Physics through the Contract No. DE-SC0012704 
and the Offices of Nuclear Physics and Advanced Scientific Computing Research within the framework of Scientific Discovery through Advance Computing 
award "Computing the Properties of Matter with Leadership Computing Resources''.

%
%
%

\end{document}